\documentclass[pra,twocolumn,showpacs]{revtex4}

\usepackage{graphicx}

\begin{document}

\title{Nonadiabatic decoherence control of qubits strongly coupled to continuum
edge}

\author{S. Pellegrin and G. Kurizki}
\affiliation{Chemical Physics Department, Weizmann Institute of Science,
Rehovot 76100, Israel}

\date{\today}

\begin{abstract}
We propose a method for contolling the decoherence of a driven qubit that is
strongly coupled to a reservoir, when the qubit resonance frequency is close to a
continuum edge of the reservoir spectrum. This strong-coupling regime is outside
the scope of existing methods of decoherence control. We demonstrate that an 
appropriate sequence of nearly abrupt changes of the 
resonance frequency can protect the qubit state from decay and decoherence more 
effectively than the intuitively obvious alternative, which is to fix the 
resonance well within a forbidden bandgap of the reservoir spectrum, as far as 
possible from the continuum edge. The "counterintuitive" nonadiabatic method 
outlined here can outperform its adiabatic counterparts in maintaining a high 
fidelity of quantum logic operations. The remarkable effectiveness of the 
proposed method, which requires much lower rates of frequency changes than
previously proposed control methods, is due to the ability of appropriately 
alternating detunings from the continuum edge to augment the interference of 
the emitted and back-scattered quanta, thereby helping to stabilize the qubit 
state against decay. Applications to the control of decoherence near the edge 
of radiative, vibrational and photoionization continua are discussed.
\end{abstract}


\maketitle

\section{INTRODUCTION}

The interaction between a spectrally structured field-continuum and a two-level 
system (TLS), which can be "dressed by coupling to near-resonant field modes, 
defines a class of fundamentally important processes in quantum field theories, 
encompassing quantum electrodynamics \cite{Cohen} and collective excitations in 
condensed media \cite{Fetter}. A diversity of peculiar features in the TLS and 
field dynamics has been shown \cite{Fain, JohnWang, Kofman, Heinzen, 
JohnTutorial, Barton} to arise when a bare or dressed TLS resonance is close to 
the continuum edge of the field spectrum. These features stem from the strong
coupling between the TLS and a continuum of field modes with an abrupt (or
nearly abrupt) cut-off and the resulting oscillatory {\em non-Markovian} 
character of decay into the continuum. Perhaps the most spectacular feature is 
the possibility of forming a {\em discrete "dressed" state}, whose energy lies
outside the continuum, in the forbidden bandgap, and is stable against decay
\cite{Fain}. The excited TLS then evolves into a superposition of a discrete,
stable state with energy in the forbidden bandgap and a decaying excited-state
component with energy in the continuum. 
The latter effect should be manifest in various systems: (i)~the partial 
inhibition of radiative decay in photonic crystals \cite{JohnWang, Kofman, 
JohnTutorial} 
and high-$Q$ cavities \cite{Heinzen, Barton, Hinds}; (ii)~the stabilization of 
a local vibrational mode against decay near the Debye cut-off of the phonon 
spectrum in a solid \cite{Kittel}; an ion trap \cite{QInfProc} or an
optical lattice \cite{Bloch}, or (iii)~the stabilization of an electronic state against
auto-ionization \cite{Rzazewski}. 
It can be interpreted as an "anomalous Lamb shift" 
\cite{JohnWang, Kofman, Heinzen, Barton} of the discrete, excited, level by the 
emitted quantum that pushes the level beyond cut-off, or as interference of the
emitted quantum with its backscattered amplitude that localizes (binds) the
excitation to the TLS. 

The present paper is dedicated to the challenge of
coherent, dynamical, control of decay and decoherence in a TLS (a qubit) whose
resonance frequency is close to a continuum edge of a reservoir, allowing for
the possibility of strong coupling between the TLS and the reservoir. We raise 
the question: how can
decoherence control benefit from the discrete, stable state of the excited TLS
in the forbidden bandgap? 
Spontaneous emission (SE) of quanta ({\it e.g.}, phonons or photons) into the
continuum of the reservoir is the dominant source of
decoherence at low temperatures, as the occupancy of the reservoir modes becomes
weak \cite{phonon, KofKurNature}. A promising means of protection from SE is to 
have the TLS resonance frequency well within the bandgap: for example, to embed
the atoms in photonic crystals (three-dimensionally periodic dielectric
structures) so as to block atomic SE at frequencies within the spectrally-wide, omnidirectional photonic bandgaps (PBGs)
\cite{PhotCrys, SEInhib, cavity}. Highly localized electronic or vibronic states
decoupled from their continua would be similarly immune to SE of their
respective quanta. 
Yet qubit manipulations for quantum information processing may necessitate TLS 
transition frequencies near the edge of the continuum, where SE is only 
{\em partially} blocked \cite{JohnWang, Kofman}. Thus, in order to coherently 
manipulate an atomic transition in the PBG via a single-photon (rather than
two-photon) transition, one must take advantage of its proximity to the edge 
frequency and
couple it to a field mode in the continuum or to a mode in the PBG created by a local
defect in the photonic crystal \cite{Kofman, Defect}
(Fig.~\ref{fig1}-Inset). Similar considerations apply to qubits that must be
intermittently coupled to vibrational continua, {\it e.g.}, trapped ions
\cite{QInfProc}, or atoms in optical lattices \cite{FortPhys}.\\
In order to 
operate quantum logic gates, based on pairwise {\em entanglement} of qubits, say 
by resonant dipole-dipole interactions \cite{dip-dipInter}, or by resonant
exchange of a phonon in an ion-trap \cite{QInfProc} or of an electron between
quantum dots \cite{Bloch} one should be able to switch the
interaction on- and off-, {\it e.g.}, by AC Stark-shifts of the transition
frequency of one qubit relative to the other. In structured continua, this would
significantly change the detuning of the resonance frequency from the continuum
edge, and thereby the degree of SE blocking. The question then arises: should
such frequency shifts be performed adiabatically? The answer is expected to be
affirmative, based on 
existing treatments of adiabatic entanglement and protection from decoherence
\cite{adiabaticpassage, ShapiroBrumer}
and on the tendency of nonadiabatic evolution to {\em spoil the gate fidelity} and promote 
transitions to the
continuum \cite{landau}. Surprisingly, the present analysis demonstrates that an appropriate
sequence of {\em "sudden"} (strongly nonadiabatic) changes of the detuning from
the continuum edge may 
yield {\em higher fidelity} of qubit and quantum gate operations than their adiabatic
counterparts. This unconventional nonadiabatic protection from decoherence can
be
attributed to the ability of appropriately alternating detunings from the
continuum 
edge in the strong-coupling regime to {\em augment the interference} of the emitted and back-scattered quanta 
amplitudes, thereby
increasing the probability amplitude of the stable state. \\
The present method is very different from previous proposals
\cite{Agarwal, Kofman-Kurizki, conclusion} to suppress
decoherence by ultra fast measurements or modulation of the coupling with the
continuum, which are all included in the {\em universal formula} of
Ref.~\cite{Kofman-Kurizki}. In order to effect decoherence suppression akin to
the quantum Zeno effect (QZE) \cite{KofKurNature, Kofman-Kurizki}, the 
measurements or modulation must be
repeated at a rate exceeding the spectra width of the continuum (its inverse
spectral width) which may be prohibitive: $10^{18}$~s$^{-1}$ for radiative 
continua \cite{KofKurNature}, and $10^{13}$~s$^{-1}$ for vibrational continua
\cite{Kittel}. By contrast, the present method requires
frequency shifts at a rate comparable to the energy of the qubit-continuum
interaction at the edge, which is typically {\em much lower} than the inverse
memory time of the continuum. 
Furthermore, the ultrafast modulation (QZE) strategy is valid only when the 
coupling to the continuum is weak,
{\it i.e.}, far enough from the continuum edge, whereas the present method works equally well for
strong or weak coupling to the continuum, since it is based on 
{\em phase-dependent} changes of the qubit state that is "dressed" by the
continuum, rather than on modulation of the "bare" state that is weakly
perturbed by the "bath" \cite{Agarwal, Kofman-Kurizki, conclusion}. 

In Sec.~\ref{HamilEqMotion} we revisit the equations of motion for partial 
decay of states near the continuum edge, allowing for the possibility of strong
coupling. In Sec.~\ref{dynamics} we introduce appropriate dynamical sequences 
of sudden frequency changes for the control of such decay. In 
Sec.~\ref{QuantGates} we discuss their implications for quantum gates and in 
Sec.~\ref{Approx} the approximations involved in our quantitative analysis. 
Experimental realizations and conclusions are discussed in
Sec.~\ref{Conclusion}. 

\section{HAMILTONIAN AND EQUATIONS OF MOTION}
\label{HamilEqMotion}

We consider a TLS with excited and ground states $\vert e \rangle$
and $\vert g \rangle$ with linear (dipolar) coupling to the field of a
discrete (or defect) mode and to a spectrally-structured mode continuum ({\it
e.g.}, radiative continuum in a photonic crystal
\cite{PhotCrys} or a high-$Q$ cavity \cite{Heinzen, Barton, Hinds}, a 
structured vibrational continuum in an ion trap or optical lattice \cite{Kittel,
QInfProc}, or an electronic continuum in a nanostructure \cite{FortPhys}). The 
hamiltonian of the system in the rotating-wave approximation assumes the form 
\cite{Kofman}

\begin{equation}
\label{hamiltonian}
\begin{array}{c}
H = \displaystyle{\hbar \omega_{at} \ \vert e \rangle \langle e \vert + \hbar \int_0^{+ \infty}
\omega \ a_\omega ^\dagger a_\omega \ \rho(\omega) \ d\omega} \\
\\
+ \ \hbar \left( \kappa_d^\star \ a_d ^\dagger \ \vert g \rangle \langle e \vert
+ h.c. \right) \\
\\
\displaystyle{+ \ \hbar \int_0^{+ \infty} \left[ \kappa^\star (\omega) 
\ a_\omega^\dagger \ 
\vert g \rangle \langle e \vert + h.c. \right] \rho(\omega) \ d\omega.}
\end{array}
\end{equation}
Here $\hbar \omega_{at}$ is the energy of the atomic transition frequency, 
$a_\omega^\dagger$ and $a_\omega$ are, respectively, the creation and annihilation 
operators of the field mode at frequency $\omega$, $\rho(\omega)$ is the mode
density of the continuum, the coupling rates between the dipole and a mode from 
the continuum or a discrete mode are $\kappa(\omega)$ and $\kappa_d$, 
respectively.\\
Let us first consider the initial state obtained by absorbing a quantum 
from the discrete mode:

\begin{equation}
\label{initcond1}
\vert \Psi (0) \rangle = \vert e,\{ 0_\omega \} \rangle,
\end{equation}
where $\vert \{ 0_\omega \} \rangle$ is the vacuum state of the field. Then the 
evolution of the wavefunction $\vert \Psi (t) \rangle$ has the form

\begin{equation}
\label{syststate}
\begin{array}{rl}
\vert \Psi (t) \rangle = & \alpha (t) \ \vert e,\{ 0_\omega \} \rangle + \beta_d
(t) \ \vert g, \ 1_d \rangle \\
\\
 & \displaystyle{+ \int_0^{+ \infty} \beta_\omega (t) \ \vert g,1_\omega \rangle \
\rho(\omega) \ d\omega}
\end{array}
\end{equation}
where we have denoted by $\vert 1_\omega
\rangle$ and $\vert 1_d \rangle$ the single-quantum 
state of the relevant modes. The Schr\"odinger equation 

\begin{equation}
i \dot \Psi (t) = H \ \Psi (t)
\end{equation}
then leads to the set of coupled differential equations

\begin{equation}
\label{systSchrod}
\begin{array}{rl}
\dot \alpha (t) = & \displaystyle{- i \ \omega_{at} \ \alpha(t) - i \ \kappa_d \ 
\beta_d (t)} \\
\\
 & - i \displaystyle{\int_0^{+ \infty} \kappa (\omega) \ \beta_\omega (t) \ 
\rho (\omega) \ d\omega}, \nonumber\\
\\
\dot \beta_d (t) = & -i \ \omega_d \ \beta_d (t) - i \ \kappa_d^\star \ \alpha
(t),\\
\\
\dot \beta_\omega (t) = & \displaystyle{-i \ \omega \ \beta_\omega (t) 
- i \ \kappa^\star (\omega) \ \alpha (t).} \nonumber
\end{array}
\end{equation}
This evolution reflects the interplay between the off-resonant Rabi oscillations
of $\vert e, \ \{0_\omega \} \rangle$ and $\vert g, \ 1_d \rangle$, at the
driving rate $\kappa_d$, and the {\em partly-inhibited oscillatory decay} from
$\vert e, \ \{ 0_\omega \} \rangle$ to $\vert g, \ \{ 1_\omega \} \rangle$ via
coupling to the continuum $\rho (\omega)$. This decay depends on the detuning of
$\omega_{at}$ from the continuum edge at $\omega_U$ (the upper cut-off of the
continuum, see Fig.~\ref{fig1}~-~inset). For a {\em spectrally steep} edge (see 
below), we are in the
regime of {\em strong coupling} to the mode continuum (as in a high-$Q$ cavity
\cite{Heinzen, Barton, Hinds, cavity})
which allows for the existence of an oscillatory, non-decaying, component of
$\alpha(t)$, associated with a discrete, stable state \cite{Kofman}.

The possibility that a stable state exists yields the inverse Laplace
transform of Eq.~(\ref{systSchrod}) and the corresponding wavefunction
(\ref{syststate}) in the form 

\begin{equation}
\label{syststatedecomp}
\vert \Psi (t) \rangle = C^{1/2} \ \vert \psi_s \rangle \ \exp(-i \ \omega_0 t) +
\vert \Psi_c (t) \rangle.
\end{equation}
Here the stable state has the energy $\hbar \omega_0$; $\vert
\psi_s \rangle$ is the stable-state eigenfunction of the Hamiltonian
(\ref{hamiltonian}) normalized to unity and weighted by the amplitude

\begin{equation}
\label{amplitude}
C^{1/2} = \left( 1 + \int_0^\infty d\omega \ \frac{\vert \kappa(\omega) \vert^2
\ \rho(\omega)}{(\omega - \omega_s)^2} \right)^{-1/2}.
\end{equation}
The explicit form of the normalized stable eigenfunction is 

\begin{equation}
\label{stablestate}
\vert \psi_0 \rangle = C^{1/2} \left( \left\vert e,\ \left\{ 0_\omega \right\}
\right\rangle - \int_0^\infty \frac{\kappa^\star (\omega)}{\omega - \omega_0} \
\left\vert g,\ 1_\omega \right\rangle \ \rho(\omega) \ d\omega \right).
\end{equation}
It is seen to be a dressed state consisting of an excited-state component and a
ground-state component, which is a superposition of contributions from all 
modes forming the "bound" quantum.\\
The other part of $\vert \Psi (t) \rangle$ in Eq.~(\ref{systSchrod}) is a
superposition of the continuous-spectrum eigenfunctions of the Hamiltonian:

\begin{equation}
\label{decaystate}
\vert \Psi_c (t) \rangle = \alpha_c (t) \ \left\vert e,\ \left\{ 0_\omega
\right\} \right\rangle + \int_0^\infty \beta_{c \omega} (t) \ \left\vert g,\
1_\omega \right\rangle \ \rho (\omega) \ d\omega,
\end{equation}
with a decaying excited-state amplitude

\begin{equation}
\label{alphaC}
\alpha_c (t) = \alpha(t) - C \ \exp (-i \ \omega_0 t) \rightarrow 0 \ \ \
\textrm{for} \ t \rightarrow \infty,
\end{equation}
and correspondingly increasing amplitudes of field modes in the continuum,

\begin{equation}
\label{betaC}
\beta_{c \omega} (t) = \beta_\omega (t) + \frac{\exp (-i \ \omega_0 t) \ C \
\kappa^\star (\omega)}{\omega - \omega_0},
\end{equation}
where $\beta_\omega (t)$ is defined by Eq.~(\ref{syststate}). \\
Since $\vert \Psi (t) \rangle$ and $\vert \psi_0 \rangle$ are all normalized to
unity, it follows that the norm of the continuum-spectrum wavefunction is

\begin{equation}
\label{NormPsiC}
\left\langle \Psi_c (t) \vert \Psi_c (t) \right\rangle = 1 - C.
\end{equation}
At $t \rightarrow \infty$ this norm becomes the probability of spontaneous
decay, which is less than unity by virtue of the existence of stable 
states in the forbidden bandgaps. When there is at most one discrete stable
state, this probability is non-zero, since $C < 1$. This allows us to conclude
that {\em there is always a non-vanishing} (albeit small) {\em probability of
decay} for an excited TLS {\em with $\omega_c$ in a forbidden band gap}.\\

Let us now introduce abrupt changes of $\omega_{at}$, {\it i.e.}, of the detuning
$\Delta_{at}=\omega_U - \omega_{at}$ from the upper cut-off, $\omega_U$, of
the continuum (by fast AC-Stark modulations as discussed below),  at intervals $\tau$.
In the sudden-change approximation for $\omega_{at}$, the amplitudes $(\alpha_{dyn}(t),\
{\beta_d}_{dyn}, \ \{ {\beta_\omega}_{dyn} (t) \})$ of the excited state, the
discrete mode and the continuum still evolve
according to Eqs.~(\ref{systSchrod}), except that from $t=0$ to $t=\tau$ the
atomic transition frequency is $\omega_{at} = \omega_A$, {\it i.e.}, the detuning
$\Delta_{at} = \omega_U - \omega_A = \Delta_A$,
while for $t> \tau $, we have $\omega_{at} = \omega_B$, {\it i.e.}, $\Delta_{at} =
\Delta_B$.
This dynamics leads to the relation

\begin{equation}
\label{dynstat1}
\begin{array}{c}
\alpha_{dyn} (t) = \alpha_A (t), \ {\beta_d}_{dyn} (t) = \beta_{d,A} (t), \\
\\
{\beta_\omega}_{dyn} (t) = \beta_{\omega,A} (t), \ (t \le \tau) \ ;\\
\\
\alpha_{dyn} (t) = \alpha_B^{(s)} (t), \ {\beta_d}_{dyn} (t) = \beta_{d,B}^{(s)}
(t), \\
\\
{\beta_\omega}_{dyn} (t) = \beta_{\omega,B}^{(s)} (t), \ (t > \tau).
\end{array}
\end{equation}
Here both $(\alpha_A (t),\ \beta_{d,A} (t), \ \{ \beta_{\omega \ A} (t) \})$ 
and $( \alpha_B^{(s)}
(t),\ \beta_{d,B}^{(s)} (t), \ \{ \beta_{\omega,B}^{(s)} (t) \})$ are solutions of
Eqs.~(\ref{systSchrod}) with a static (fixed) atomic transition frequency, $\omega_A$
or $\omega_B$. However, the initial condition at the instant $t = \tau$ of the 
frequency change from $\Delta_A$ to $\Delta_B$ is no longer the excited
state (\ref{initcond1}) but the superposition:

\begin{equation}
\label{initcond2}
\begin{array}{rl}
\vert \Psi (\tau) \rangle = & \alpha_A (\tau) \ \vert e,\ \{ 0_\omega \} \rangle +
\beta_{d,A}(\tau) \ \vert g, \ 1_d \rangle \\
\\
 & \displaystyle{+ \int_0^{+ \infty} \beta_{\omega,A} 
(\tau) \ \vert g,\ 1_\omega \rangle \ \rho (\omega) \ d\omega.}
\end{array}
\end{equation}
In other words, the dynamics is equivalent to two successive static evolutions,
the second one starting from initial conditions $(\alpha_A (\tau),\ \beta_{d,A}
(\tau), \ \{
\beta_{\omega,A} (\tau) \})$. \\
Using the Laplace transform of the
system~(\ref{systSchrod}) with the initial condition~(\ref{initcond2}), it is
possible to express the dynamic amplitude of the excited state after the sudden 
change as

\begin{equation}
\label{dynstat2}
\begin{array}{c}
\alpha_{dyn} (t)= \alpha_A (\tau) \ \alpha_B (t-\tau) + \beta_{d,A} (\tau) \
\beta_{d,B} (t-\tau) \\
\\
\displaystyle{+ \int_0^{+ \infty} \beta_{\omega,A} (\tau) \ 
\beta_{\omega,B} (t-\tau) \ \rho(\omega) \ d\omega, \ (t>\tau),}
\end{array}
\end{equation}
where we have used the initial conditions $(\alpha_A(\tau),\
\beta_{d,A} (\tau), \ \{
\beta_{\omega,A} (\tau) \})$ and the solution $(\alpha_B(t), \ \beta_{d,B} (t),
\ \{
\beta_{\omega,B} (t) \})$ of Eqs.~(\ref{systSchrod}) for the initial 
condition~(\ref{initcond1}). 

\section{INTERFERING SUCCESSIVE EVOLUTIONS: "COUNTERINTUITIVE" SEQUENCE}
\label{dynamics}

Clearly, Eq.~(\ref{dynstat2}) is sensitive to the
{\em relative phases} of the successive static evolutions (labelled by A and B),
{\it i.e.},
to their {\em interference}. The contribution of the first term in 
(\ref{dynstat2}) to the excited-state population 
$\vert \alpha_{dyn}(t) \vert^2$ always decreases after the sudden change and 
then oscillates, before settling to an asymptotic non-zero value. On the other 
hand, the contribution of the second and third terms and their cross-product 
with the first term increases immediately after the sudden change. Yet whatever 
the time $\tau$ of the sudden change, when performing only one change, the 
increasing contribution is never large enough to compensate for the decreasing 
part. Then the dynamic population of the excited state after the sudden change 
always lies inbetween the two static populations obtained for $\Delta_A$ and 
$\Delta_B$.

\begin{figure}
\begin{center}
\includegraphics[width=9cm]{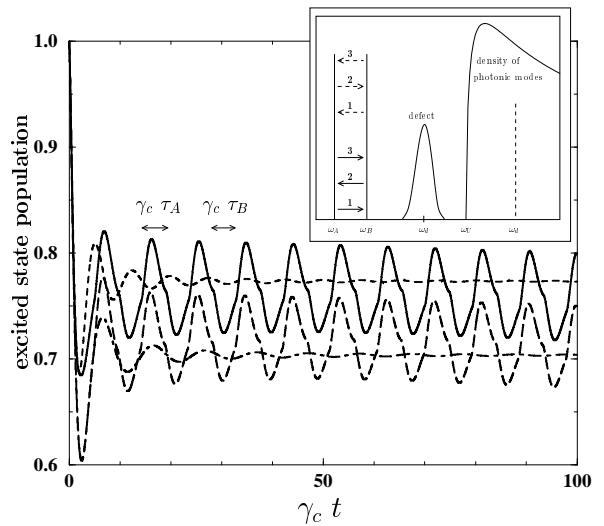}
\caption{Excited state population as a function of dimensionless time $\gamma_c \ t$.
Dashed line: static detuning $\Delta_A / \gamma_c = 0.5$. Dot-dashed line:
static detuning $\Delta_B / \gamma_c = 0.25$. Solid and long-dashed lines:
periodic sudden
shifts of the detuning between $\Delta_A$ and $\Delta_B$. Solid line: starting 
with detuning $\Delta_A$ (solid arrows in inset).
Long-dashed line: starting with detuning $\Delta_B$ (dashed arrows in inset). 
Frequencies $\omega_A$ and $\omega_B$ are in the
vicinity of a discrete mode $\omega_d$ and of a PBG edge $\omega_U$ (inset)}
\label{fig1}
\end{center}
\end{figure}

There is, however, an advantageous feature to the sudden change: since the time 
dependence of
$\alpha_{dyn}(t)$ in (\ref{dynstat2}) arises from the static amplitudes $\alpha_B$, $\beta_{d,B}$ and 
$\beta_{\omega,B}$
at the {\em shifted} time $t-\tau$, a consequence of the sudden change is to
revive the excited-state population oscillations, which tend to disappear
at long times in the static case. Hence, by applying several {\em successive} 
sudden changes,
we should be able to maintain large-amplitude oscillations of the {\em
coherence} between $\vert e \rangle$ and $\vert g \rangle$. The scenario leading
to the largest amplitude consists in {\em periodic} shifts of the energy detuning from
$\Delta_A$ to $\Delta_B$. Here we have the choice between starting from
$\Delta_{at} = \Delta_A$
or $\Delta_{at} = \Delta_B$ (Fig.~\ref{fig1}-inset). Analysis of
Eq.~(\ref{dynstat2}) then shows that in the former case, the dynamic 
population experiences large amplitude oscillations but never exceeds the highest 
static population. But when the initial detuning $\Delta_A$ is large and we first 
reduce it to $\Delta_B$ before it increases to $\Delta_A$, the {\em dynamic
population and the $\vert e \rangle - \vert g \rangle$ coherence}, thanks to the 
revival of oscillations, are {\em periodically larger} than
the static ones (!).  This is illustrated in Fig.~\ref{fig1} (see
Sec.~\ref{Approx} for quantitative assumptions).

This remarkable result comes about unexpectedly, since it implies that 
successive abrupt changes can {\em reverse} the
decay to the continuum, even though they {\em cannot be associated with the
quantum Zeno effect} \cite{Rzazewski, KofKurNature} or with its ultrafast
modulation counterparts \cite{Agarwal, Kofman-Kurizki, conclusion}: 
the present abrupt changes occur at intervals much longer than the correlation 
(Zeno) time of the continuum, which is utterly negligible for the case of the
radiative (electromagnetic) continuum ($10^{-18}$~s) \cite{KofKurNature}. The
required intervals are even longer than the static-oscillation half-period.
The fact that this happens only for the rather "counter-intuitive" ordering of
detuning values (from large to small then back again) is a manifestation of
interference between successive static evolutions: their relative phases
determine the beating between the emitted and reabsorbed (back-scattered) photon
amplitudes and thereby the oscillation of $\alpha_{dyn}(t)$.\\
Let us now consider the initial superposition
\begin{equation}
\label{initcond3}
\vert \Psi (0) \rangle = \alpha (0) \ \vert e,\{ 0_\omega \} \rangle + \beta_d (0)
\ \vert g,1_{\omega_d} \rangle
\end{equation}
and a non-negligible coupling constant $\kappa_d$.
In this case, the periodic dynamic population of the excited state also
strongly
exceeds the static one. On the other hand, the discrete mode amplitude 
$\beta_d(t)$ diminishes as compared with the static case. Most importantly, the 
{\em instantaneous dynamic fidelity} $\vert \langle \Psi (0) \vert \Psi (t) \rangle 
\vert^2$ is periodically enhanced as compared to the static one, as demonstrated
numerically (dot-dashed line) in Fig.~\ref{fig2} (see Sec.~\ref{Approx} for
quantitative assumptions).

\section{APPLICATION TO CONTROL-PHASE GATES}
\label{QuantGates}

\begin{figure}
\begin{center}
\includegraphics[width=9.0cm]{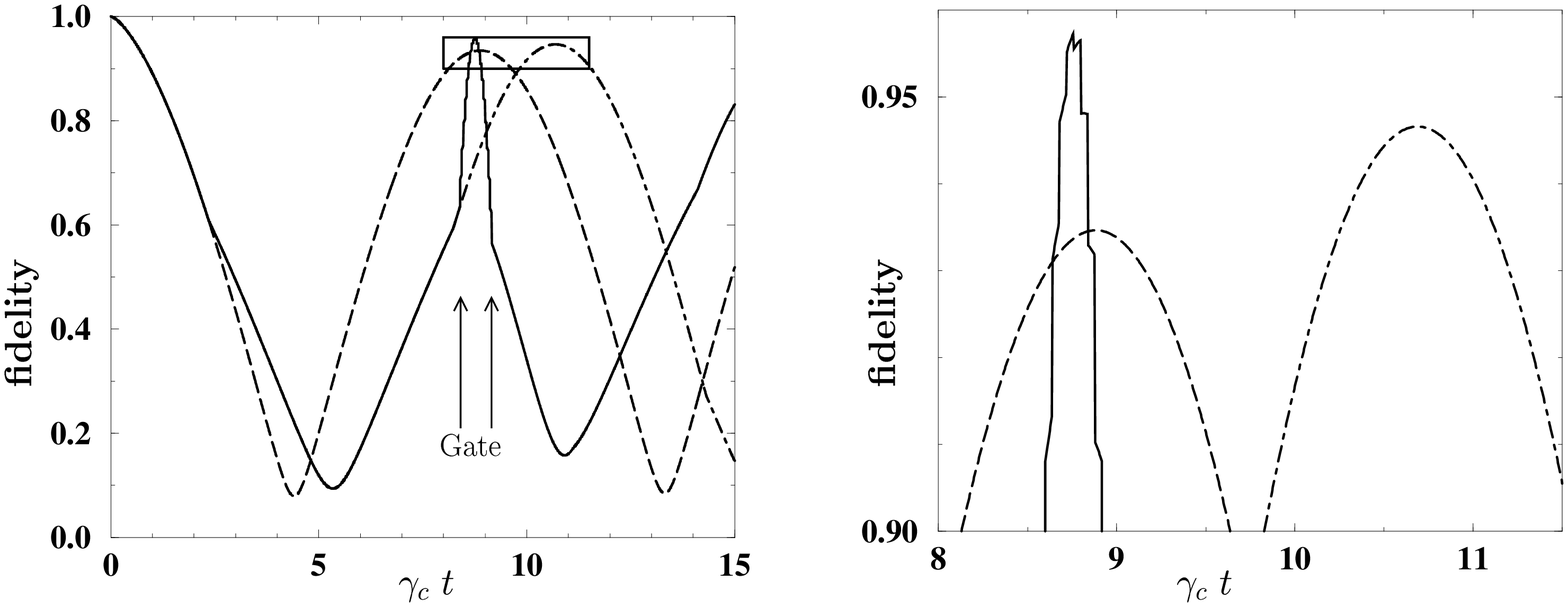}
\caption{Fidelity of a superposition state (\ref{initcond3}) as a function 
of the dimensionless time $\gamma_c t$. Long-dashed line: static detuning
$\Delta_A / \gamma_c = 0.5$. Dot-dashed line: periodic sudden shifts of the 
detuning from $\Delta_A / \gamma_c = 0.5$ to $\Delta_B / \gamma_c =
0.25$. Solid line: periodic detuning shifts alternating with
control phase shifts, effecting the control phase gate. Arrows mark the start 
and the end of the gate operation. Right panel: enlarged view of the rectangle.}
\label{fig2}
\end{center}
\end{figure}

In order to use these results for quantum logic
gates, let us consider the example of control-phase 
gate, which consists in 
shifting the phase of the target-qubit excited state by $\pi$ via interaction with
the control qubit \cite{dip-dipInter}. Such a gate is characterized by the truth
table \cite{NielsenChuang}

\begin{equation}
\begin{array}{l}
\vert 0 \rangle \ \vert 0 \rangle \rightarrow \vert 0 \rangle \ \vert 0 \rangle,
\\
\\
\vert 0 \rangle \ \vert 1 \rangle \rightarrow \vert 0 \rangle \ \vert 1 \rangle,
\\
\\
\vert 1 \rangle \ \vert 0 \rangle \rightarrow \vert 1 \rangle \ \vert 0 \rangle,
\\
\\
\vert 1 \rangle \ \vert 1 \rangle \rightarrow e^{i \pi} \ \vert 1 \rangle \
\vert 1 \rangle.
\end{array}
\end{equation}
In order to ensure compatibility with the proposed decoherence control, the 
phase shift must be accumulated gradually, so as to preserve the coherence of 
the system. We have found that a single sudden shift of $\pi$ is incompatible
with our method. By contrast, ten or twenty sudden shifts of $\pi / 10$ or 
$\pi / 20$, respectively, alternating with an appropriate sequence of detuning
changes, keep the fidelity high, with little decoherence. 
Without attempting to fully optimize the process, we have been
able to find such dynamics of the shift that preserves a high fidelity of the
system state. The system begins to evolve
following the "counter-intuitive" detuning sequence discussed in
Sec.~\ref{dynamics} (not to be confused with the adiabatic STIRAP method 
\cite{adiabaticpassage}!).
As soon as two sudden changes of the detuning have been performed, the
conditional phase shifts of $\pi / 10$ or $\pi / 20$ take place, to be followed
by two more sudden changes and so forth, each time optimizing the detuning to
obtain the best protection against spontaneous emission. The total gate 
operation is completed within the time-interval of maximum fidelity as seen in
Fig.~\ref{fig2}. 

Figure \ref{fig2}, showing the fidelity of the system relative to
its initial state during the realization of a control phase gate according to
the procedure described above, is perhaps our most impressive finding. We can 
see that the fidelity is increased using the "counterintuitive" sequence of
detunings (solid line) as compared to the static (fixed) choice of maximal 
detuning (long-dashed line), or compared to the dynamically enhanced fidelity 
$\vert \langle \Psi (0) \vert \Psi (t) \rangle \vert^2$ obtained without 
gate operations (dot-dashed line). \\

\section{APPROXIMATIONS AND NUMERICAL SOLUTIONS}
\label{Approx}

While the results in Fig.~\ref{fig1} and \ref{fig2} are generic, they have been
obtained based on certain quantitative assumptions. In what follows we outline
these assumptions and their numerical implementation. \\

\subsection{The effective mass approximation and continuum discretization}

In our numerical studies, we have applied the results of the foregoing general analysis to a
model density-of-modes (DOM) distribution. This distribution is derived on keeping the
lowest term in the Taylor expansion of the dispersion relation $\omega (\vec k)$
near a cut-off frequency $\omega_U$ in a 3D-periodic structure (photonic,
electronic or vibronic crystal). This yields

\begin{equation}
\label{dispersion}
\omega \approx \omega_U + \sum_{i=x,\ y,\ z} A_i \ (k - k_U)_i^2,
\end{equation}
which is known as the effective-mass approximation \cite{Pantelides}. In a structure
with period $L$, $k_U$ satisfies the Bragg condition $k_U = \pi / L$. The
corresponding DOM in a three-dimensionally periodic structure with an allowed
point-group symmetry may be approximated as \cite{JohnWang}

\begin{equation}
\label{density}
\rho (\omega) \propto (\omega - \omega_U)^{(1-D)/2} \ \theta (\omega - \omega_U),
\end{equation}
where $\theta$ is the step function and $D$ is the dimension of the
Brillouin-zone surface spanned by band-edge modes with vanishing group velocity.
All parameters and variables in Figs.~\ref{fig1} and \ref{fig2} are scaled to 
the effective coupling $\gamma_c=\kappa^2 / \sqrt{\epsilon}$, which depends on 
the edge steepness $\epsilon$. 

Results discussed in Secs.~\ref{dynamics} and \ref{QuantGates} have been 
obtained in the isotropic dispersion approximation, corresponding to $D=2$
\cite{JohnTutorial,Kofman}. In the domain of photonic crystals such an 
approximation yields qualitatively correct results \cite{JohnTutorial,
PhotCrys}. 

\begin{figure}
\begin{center}
\includegraphics[width=8.0cm]{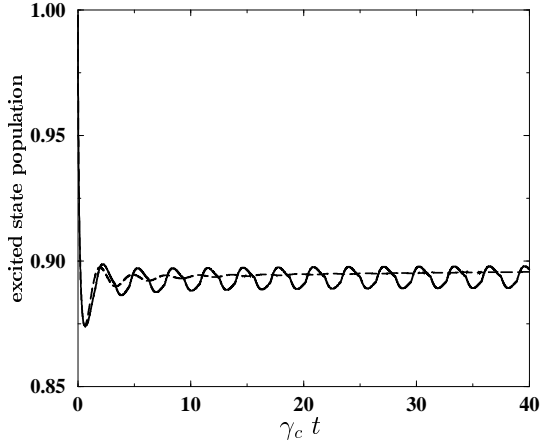}
\caption{Excited state population as a function of dimensionless time $\gamma_c
\ t$ in the case of an anisotropic density of modes. Initial and final detunings
are the same as in Fig.~\ref{fig1}: $\Delta_A / \gamma_c = 0.5$ and $\Delta_B /
\gamma_c = 0.25$. Long-dashed line: static case for detuning $\Delta_A$. Solid
line: periodic counterintuitive dynamics.}
\label{figure3}
\end{center}
\end{figure}

The modification of these results to
allow for the anisotropy of the density of modes in a realistic photonic crystal
can be undertaken using the {\em anisotropic dispersion relation} with $D=0$. 
This leads to a DOM \cite{JohnTutorial, PhotCrys} 

\begin{equation}
\label{densityAnisotrop}
\rho (\omega) \sim (\omega - \omega_U)^{1/2}.
\end{equation}
Figure \ref{figure3} shows that our results concerning the efficiency of the periodic
counterintuitive dynamics remains true when considering the anisotropy of the
periodic structure. In order to numerically solve the system (\ref{systSchrod}), 
the continuum has been
discretized. The coupling $\kappa(\omega)$ has been approximated to be constant
(independent of $\omega$) and real, and the analytic expression (\ref{density})
or (\ref{densityAnisotrop}) for the DOM has been used. 

\subsection{Validity of the dynamical analysis}

\begin{figure}
\begin{center}
\includegraphics[width=8.0cm]{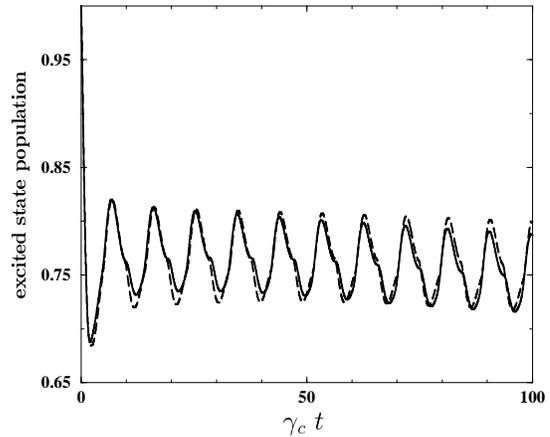}
\caption{Influence of finite transition times as compared with the sudden
change approximation. Long-dashed line: periodic counterintuitive dynamics in
the sudden change approximation (same curve as solid line of Fig.~\ref{fig1}).
Solid line: same periodic counterintuitive dynamics with finite transition
times. Initial and final detunings are $\Delta_A / \gamma_c = 0.5$ and 
$\Delta_B / \gamma_c = 0.25$.}
\end{center}
\label{fig4}
\end{figure}

Since the sudden change approximation is not realizable experimentally, we
have considered the effects of finite transition times between $\Delta_A$ and
$\Delta_B$, by using a sequence of pulses that vary as $\exp [-(t-\tau_n)^8/\tau^8]$,
{\it i.e.}, are centered on $\tau_n$ and have half-width $\tau$ of the order of 
$\tau_A / 4 \sim \tau_B /4$. The excited state population is only slightly 
modified by such finite rise- and falloff-times (Fig.~\ref{fig4}). The
rotating-wave approximation is obeyed, since the switching time is long enough: 
$\tau \gg 1/\omega_0$. \\
We have compared our results, which allow for possibly strong 
coupling of $\vert e \rangle$ with the continuum edge, with those of the 
universal formula of 
Ref.~\cite{Kofman-Kurizki}. This formula expresses the decay rate of $\alpha
(t)$ by the convolution of the periodic modulation spectrum and the continuum 
coupling spectrum. We find good agreement with this formula only in the regime
of {\em weak coupling} to the continuum edge, when the dimensionless detuning parameter
$\Delta_{at} / \gamma_c > 5$, as expected from the limitations of the theory in 
Ref.~\cite{Kofman-Kurizki}. We note that the phase-modulation models of
Refs.~\cite{Agarwal} or \cite{conclusion} are not applicable to the present 
situation.\\

\section{DISCUSSION}
\label{Conclusion}

The following experimental scenarios may be envisioned for demonstrating the 
proposed control of decoherence. (a)~Pairs of qubits are realizable by two
species of active rare-earth dopants \cite{dopant} or quantum dots in a photonic
crystal. 
(b)~Alternatively these may be atoms within a high-$Q$ cavity that may have
different, {\em controllable} coupling strengths to the cavity mode. (c)~Another
candidate system is that of neighboring atoms or ions trapped in optical
lattices or ion-arrays when their local vibrational resonances are close to the
Debye cutoff of the structure.

It is possible to control the rate of spontaneous emission and the frequency of
excited-state oscillations by varying only the two detuning parameters
$\Delta_A$ and $\Delta_B$. 
Let us assume that the transition 
frequency of one qubit is initially detuned by $\Delta_A \sim 4$~MHz from 
the continuum edge with coupling constant $\gamma_c \sim 10$~MHz, and by $\sim
3$~MHz  
from the resonance of the other qubit. This $\Delta_A$ is abruptly 
modulated by fsec non-resonant laser pulses, which exert $\sim 3$~MHz AC Stark 
shifts. Between successive shifts, the neighboring qubits are near-resonant with
each other and therefore can become entangled ({\it e.g.}, dipole-dipole
coupled), thus effecting the
high-fidelity phase-control gate operation \cite{dip-dipInter}, as in
Fig.~\ref{fig2}. The required pulse rate is $\gamma_c/10~\sim 1$~MHz, much lower
than the pulse rate stipulated under similar conditions by the previously 
proposed ultrafast-modulation~/~quantum-Zeno strategies
\cite{Agarwal,Kofman-Kurizki, conclusion}.

To summarize, we have discovered that a ''counterintuitively'' ordered sequence of
abrupt changes of the detuning between the qubit transition and a continuum edge
(the photonic band cut-off), is able to protect
the qubit state from spontaneous emission of photons or phonons more effectively than the intuitively
obvious alternative, which is to fix the 
largest possible detuning value. This method is effective even under conditions
of strong coupling to the continuum , as opposed to previously proposed phase-
or frequency-modulation strategies
\cite{KofKurNature,Agarwal,Kofman-Kurizki, conclusion}. The present method is a highly advantageous 
means of maintaining high fidelity of quantum
states and quantum-logic operations in the presence of decoherence by
nonadiabatic interference, contrary to 
prevailing adiabatic approaches to quantum-state control. This may pave the way 
to new methods of controlling decay and decoherence in spectrally structured
continua.

\acknowledgments 
We acknowledge the support of the EC Human 
Potential Programme (HPRN-CT-2002-00309, QUACS), ISF and Minerva.

\end{document}